\documentclass[12pt]{article}
\usepackage{amsmath}
\usepackage{amssymb}
\usepackage{cancel}
\usepackage{graphicx}

\numberwithin{equation}{section}

\begin{document}
\baselineskip=19pt

\begin{titlepage}
\begin{flushright}
{\small KUNS-2278}
\end{flushright}

\begin{center}
\vspace*{17mm}

{\large\bf%
Seesaw in the bulk
}

\vspace*{10mm}

Atsushi Watanabe,$^*$ 
~Koichi Yoshioka$^\dag$
\vspace*{2mm}

$^*${\small {\it Department of Physics, 
Niigata University, Niigata 950-2181, Japan}},\\
$^\dag${\small {\it Department of Physics, Kyoto University, Kyoto 
606-8502, Japan}}\\

\vspace*{3mm}

{\small (July, 2010)}
\end{center}

\vspace*{7mm}

\begin{abstract}\noindent%
A five-dimensional seesaw framework is analyzed with the
lepton-number-violating propagator of bulk right-handed
neutrinos. That can bypass summing up the effects of heavy Majorana
particles whose masses and wavefunctions are not exactly known. The
propagator method makes it easier to evaluate the seesaw-induced
neutrino mass for various boundary conditions of bulk neutrinos and in
a general background geometry, including the warped extra dimension. It
is also found that the higher-dimensional seesaw gives a natural
framework for the inverse seesaw suppression of low-energy neutrino
masses.
\end{abstract} 

\end{titlepage}

\newpage

\section{Introduction}
The discovery of nonzero neutrino masses is one of the most impressive 
developments in particle physics recently made. In addition to
cosmological observations, the flux measurements of solar and
atmospheric neutrinos indicate that neutrino masses are tiny compared
to the other fermion masses~\cite{nu}. The smallness of neutrino masses
may be regarded as an indication of higher energy scale than the
electroweak one, possibly connected with deeper concepts such as
grand unification and other anticipated scenarios beyond the Standard
Model (SM).

As a feasible paradigm to address problems in the SM, theories with
extra spatial dimensions have been widely studied over the past
decade. For example, the gauge hierarchy problem is solved by large
volume of the extra space which makes the Planck scale suppressed down
to TeV~\cite{ADD}. The localized gravity with the warped
metric~\cite{RS} also provides a possible interpretation of the gauge
hierarchy by small overlap between matter and gravitational
fields. Interestingly, these mechanisms for the hierarchy problem also
control the neutrino mass. As in the same way that the gravitational
flux is diluted, the neutrino mass is suppressed if gauge-singlet
neutrinos propagate in the bulk~\cite{bulk_nuR}. For the warped
extra dimension, the localization of bulk neutrinos produces tiny
Dirac neutrino masses~\cite{warp_nuR}. Thus the connection between
neutrino physics and extra dimensions has been a subject of great
interests to particle physics~\cite{5d_nu}.

Motivated by this phenomenological connection, we discuss the seesaw
mechanism in higher-dimensional theory where right-handed Majorana
neutrinos spread over the extra space. Our main emphasis in this paper
is on the propagator for bulk Majorana fermion which serves as an
useful tool to calculate the low-energy neutrino mass induced by the
seesaw operation. In Section 2, we discuss the setup of the
higher-dimensional seesaw and formulate the bulk field propagator. In
Section 3, we present various applications of the propagator method to
higher-dimensional seesaw models. Section 4 is devoted to summarizing
the results. Appendices~\ref{LG} and \ref{PVBC} show our convention
for Lorentz spinors and the derivation of propagators in various
situations.

\medskip

\section{Seesaw in five dimensions}
In this section, we introduce the framework of higher-dimensional
seesaw mechanism and illustrate the Kaluza-Klein (KK) expansion and
the propagator method to obtain low-energy neutrino masses.

\begin{figure}[t]
\begin{center}
\scalebox{0.43}{\includegraphics{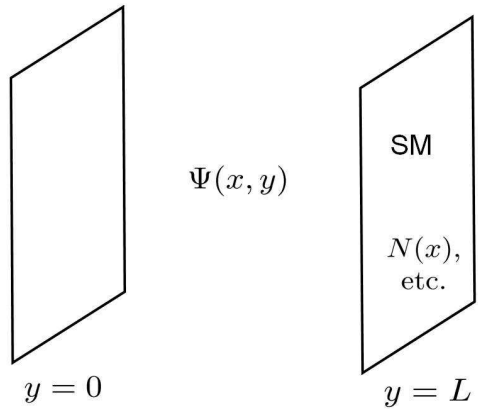}}
\end{center}
\caption{A sketch of the model. The SM fields are localized 
at $y=L$ while the right-handed neutrinos $\Psi(x,y)$ spread over the
extra space with bulk Dirac and Majorana masses.\medskip}
\label{fig1}
\end{figure}

\subsection{Setup}
\label{base}
The fifth dimension $y$ is compactified on the $S^1/Z_2$ orbifold such
that there are two fixed points at $y =0$ and $L$. The four-component
bulk fermions $\Psi(x,y)$ are introduced as right-handed neutrinos and
obey the boundary conditions associated with two operations on 
the $S^1/Z_2$ space; the translation $\hat{T} : y \to y + 2L$ and the 
parity $\hat{Z}: y \to -y$. These conditions are written as
\begin{eqnarray}
\Psi(x,-y) = (Z\otimes\gamma_5) \Psi(x,y), \quad\quad 
\Psi(x,y + 2L) = T \, \Psi(x,y), 
\label{zt}
\end{eqnarray}
where $Z$ and $T$ are the matrices acting on the field space. The
translation and parity imply that $Z^2 = 1$ and $ZT = T^{-1}Z$ should
be satisfied. Instead of the translation $T$ in (\ref{zt}), another
parity $Z'=TZ$ can be used to define the boundary conditions; 
\begin{eqnarray}
\Psi(x,-y) = (Z\otimes\gamma_5) \Psi(x,y), \qquad
\Psi(x,L - y) = (Z'\otimes\gamma_5) \Psi(x,L + y). 
\label{zzp}
\end{eqnarray}
The parity $Z'$ is the reflection with respect to $y = L$ and the
equations (\ref{zzp}) choose the Dirichlet or Neumann condition at
each boundary. In this section, we consider the standard 
condition $Z = 1$ and $Z' = 1$ with which the upper (right-handed)
component has the zero mode. The other possibilities and their
physical implication will be discussed in Section~\ref{appli} and
Appendix~\ref{PVBC}.

The SM fields including the left-handed neutrinos $N= P_L N =
\big(\begin{smallmatrix}0 \\ \nu_L^{}\end{smallmatrix}\big)$ and the
Higgs field $H$ are assumed to be localized 
at $y = L\,$ (Fig.~\ref{fig1}). This SM-field profile gives an example
and the analysis below is applied to other cases in a similar
manner. The seesaw mechanism in five dimensions is described by the
following bulk and boundary Lagrangians; 
\begin{eqnarray}
\mathcal{L}_{\rm bulk} &=&
i\overline{\Psi} \Gamma^M \partial_M \Psi
- m_d\theta(y)\overline{\Psi}  \Psi 
- \frac{1}{2} \big(M \overline{\Psi^c}\Psi + {\rm h.c.}\big),
\label{Lbulk} \\
\mathcal{L}_{\rm bound} &=&
-\left(\, \frac{m}{\sqrt{\Lambda}} \overline{\Psi}N  
\,+\, {\rm h.c.}\right) \delta(y - L),
\label{Lbound}
\end{eqnarray}
where $\Lambda$ stands for the fundamental scale of the theory. We
have introduced the bulk Dirac mass $m_d$ with the step 
function $\theta(y)$ which is needed to implement the $Z_2$
invariance. The mass parameters $m_d$ and $M$ are assumed to be flavor
diagonal in this section, while that will be relaxed later. The
boundary Dirac mass $m$ is made out of the neutrino Yukawa coupling
and the vacuum expectation value of the Higgs 
field $\langle H \rangle = 
\big(\begin{smallmatrix}0 \\ v\end{smallmatrix}\big)$. The
charge-conjugate spinor $\Psi^c$ is defined 
by $\Psi^c \equiv \Gamma^3\Gamma^1 \overline{\Psi}^{\rm T}$. Our
convention for the gamma matrices and Lorentz spinors are given in
Appendix \ref{LG}.

\medskip

\subsection{The KK expansion}
\label{KK}
There are two physically equivalent, but different prescriptions to
derive four-dimensional effective theory from the original
five-dimensional Lagrangian: the KK expansion and the propagator
method for bulk fields.

With the KK expansion, the neutrino spectrum is obtained by the
diagonalization of the infinite-dimensional mass matrix spanned by KK
modes and the SM neutrinos. A bulk field $\Psi(x,y)$ is expanded as
\begin{equation}
\Psi(x,y) \;=\; \left(\begin{array}{c}
\sum_n\chi_R^n(y) \psi_R^n(x) \\[2mm]
\sum_n\chi_L^n(y) \psi_L^n(x)
\end{array}\right)
\end{equation}
with the orthogonal systems $\chi_{R,L}^n(y)$. It is convenient to
choose them to satisfy
\begin{eqnarray}
\big[ \partial_y + m_d\theta(y) \big]\chi_R^n &=& +M_{K_n}\chi_L^n, 
\\[.5mm]
\big[ \partial_y - m_d\theta(y) \big]\chi_L^n &=& -M_{K_n}\chi_R^n, 
\end{eqnarray}
and the normalization conditions $\int_0^L dy \,\,\chi^m_{R,L}
\!\!{}^\dag \, \chi^n_{R,L} = \delta_{mn}$. Under the boundary
conditions $Z = 1$ and $Z' =1$, the expansion 
functions $\chi_{R,L}^n$ and the KK mass $M_{K_n}$ are found to be
\begin{eqnarray}
&&\chi_R^0 \,=\,  \sqrt{\frac{2}{L}} N_0 \, e^{-m_d |y|},\\
&&\chi_R^n \,=\,  \sqrt{\frac{2}{L}} \left[
-N_n \cos\left( \frac{n\pi}{L}y \right) + 
\sqrt{1 - N_n^2} \,\theta(y)\sin\left( \frac{n\pi}{L}y \right) 
\right]  \quad (n \geq 1),\\
&&\chi_L^n \,=\,  \sqrt{\frac{2}{L}}
\sin\left( \frac{n\pi}{L}y \right) \qquad (n \geq 1),\\[3mm]
&& \hspace{25mm}
M_{K_n} \, = \, \sqrt{m_d^2 + \left(\frac{n\pi}{L}\right)^2 }
\qquad (n \geq 1),
\end{eqnarray}
where $N_0 = \sqrt{m_dL/(1 - e^{-2m_d L})}$
and $N_n = \left(\frac{n\pi}{L}\right)/M_{K_n}$. The zero-mode
function $\chi_R^0$ is localized at $y=0$ ($y=L$) due to the bulk
Dirac mass if $m_d > 0$ ($m_d <0$).

By substituting the KK expansion into the five-dimensional Lagrangian
and integrating it over the extra space, we have the four-dimensional
effective Lagrangian
\begin{eqnarray}
\mathcal{L}_4 \,=\, 
i \mathcal{N}^\dag  \sigma^\mu \partial_\mu \mathcal{N}
-\frac{1}{2}\left( \mathcal{N}^{\rm T} \epsilon \mathcal{M} \mathcal{N}
+ {\rm h.c.} \right),
\end{eqnarray}
\begin{eqnarray}
\mathcal{M} =\!
\left( \begin{array}{c|c}
 0 &  \,\, \mathcal{M}_D^{\rm T} \, \\[1mm]\hline
 &  \\
\!\!\mathcal{M}_D\! &  \;\; \mathcal{M}_H  \, \\
 &  \\
\end{array}\right)
= 
\left( \begin{array}{c|cccc}
 & m_0^{\rm T} & m_1^{\rm T} &  & \cdots \\\hline
\!m_0 & -M_{R_{00}}^* &  &  & \\
\!m_1 &  & \!\!-M_{R_{11}}^* & \!\!M_{K_{1}} &  \\
 &  & \!\!\!\phantom{-}M_{K_{1}} & \!\!M_{L_{11}} &  \\
\vdots &  &  &  & \ddots \\
\end{array}
\right)\!\!,\quad 
\mathcal{N} =\!
\begin{pmatrix}
\nu_L^{} \\[.5mm]  \epsilon\psi_R^0{}^* \\[.5mm]
\epsilon \psi_R^1{}^* \\[.5mm]  \psi_L^1 \\
\vdots \\
\end{pmatrix}\!\!,\;\;
\label{KKmm}
\end{eqnarray}
\begin{gather}
M_{R_{nm}} \,=\,  
\int_{0}^{L} \!\! dy \,\,  \chi_R^n{}^{\rm T}M \chi_R^m 
\,=\, M\delta_{nm},  \qquad
M_{L_{nm}} \,=\, 
\int_{0}^{L} \!\! dy \,\,  \chi_L^n{}^{\rm T}M \chi_L^m 
\,=\, M\delta_{nm},\nonumber\\[1mm]
m_n \,=\, \frac{m}{\sqrt{\Lambda}}
 \chi_R^n{}^\dag(L) = 
\left\{ \begin{array}{ll}
m\sqrt{\frac{2}{\Lambda L}}N_0\,e^{-m_d L} & (n = 0), \\[2mm]
m\sqrt{\frac{2}{\Lambda L}}N_n (-1)^{n+1} & (n \geq 1).
\end{array}\right.
\label{KKmm2}
\end{gather}
The mass spectrum of Majorana neutrinos is obtained by 
diagonalizing $\mathcal{M}$. For $\mathcal{M}_D \ll \mathcal{M}_H$,
the seesaw mechanism works and the Majorana mass of light 
neutrinos $M_\nu$ is approximately given by
\begin{eqnarray}
M_\nu \,=\, - \mathcal{M}_D^{\rm T} \mathcal{M}_H^{-1} \mathcal{M}_D.
\label{seesawKK}
\end{eqnarray}
The KK expansion describes the seesaw-induced mass $M_\nu$ as the
summation of KK neutrino contributions; 
\begin{eqnarray}
M_\nu \!&=& 
-\begin{pmatrix} m_0^{\rm T} \;\; m_1^{\rm T} \;\; 0 \;\; \cdots
\end{pmatrix}
\begin{pmatrix}
\frac{-1}{M^*} & & & \\
& \!\frac{-M}{|M|^2 + M_{K_1}^2} 
& \frac{M_{K_1}}{|M|^2 + M_{K_1}^2} &  \\
& \!\frac{M_{K_1}}{|M|^2 + M_{K_1}^2} 
& \frac{M^*}{|M|^2 + M_{K_1}^2} &  \\
&  &  & \ddots \\
\end{pmatrix}
\begin{pmatrix}
m_0 \\[1.3mm]
m_1 \\[1.7mm]
0 \\[1.2mm]
\vdots
\end{pmatrix}\nonumber\\
&=& \frac{2}{\Lambda L} \left(
\frac{m_d L\,e^{-2m_d L}}{1 - e^{-2m_d L}} \,+ \,
\sum_{n=1}^{\infty} \frac{(n\pi)^2}{(m_d L)^2 + (n\pi)^2}\cdot
\frac{|ML|^2}{|ML|^2 + (m_dL)^2 + (n\pi)^2 }\right)
\frac{m^{\rm T}m}{M^*}
\nonumber\\[1mm]
&=& \frac{1}{\Lambda L}\,
\frac{\widetilde{M}L \cosh(\widetilde{M}L)-m_d L\sinh(\widetilde{M}L)}
{ \sinh(\widetilde{M}L)}\,
\frac{m^{\rm T}m}{M^*},
\label{KKsum}
\end{eqnarray}
where $\widetilde{M} = \sqrt{m_d^2 + |M|^2}$. We will discuss physical
implications of this result in Section~\ref{appli}.

\medskip

\subsection{Propagators for bulk Majorana fermions}
\label{prop}
In the propagator method, the low-energy neutrino mass is calculated
by treating the bulk-boundary mixing (\ref{Lbound}) as a small 
perturbation. Let us start with the functional integral
\begin{eqnarray}
Z &=& \int\!\mathcal{D}\Psi^* \mathcal{D}\Psi
\,\exp \bigg[ i \! \int \! d^5x\,
(\mathcal{L}_{\rm bulk} + \mathcal{L}_{\rm bound} ) \bigg]
\nonumber\\[1mm]
&=&
\int\! \mathcal{D}\Psi^* \mathcal{D}\Psi
\bigg[\,1 + \frac{1}{2}\bigg(i\!\int \! d^5x\,\mathcal{L}_{\rm bound} 
\bigg)^2 + \cdots \bigg]
\exp \bigg( i \! \int \! d^5x\, \mathcal{L}_{\rm bulk} \bigg).
\end{eqnarray}
After integrating out the bulk fermions $\Psi(x,y) =
\left(\begin{smallmatrix}\xi(x,y) \\ 
\eta(x,y)^{}\end{smallmatrix}\right)$,
the lepton-number-violating part in the quadratic term is given by
\begin{eqnarray}
\frac{(i)^2}{2} \!\! \int \! d^4x d^4x' 
\left[ 
-\overline{N^c}(x) \frac{m^{\rm T}}{\sqrt{\Lambda}} 
\int \!\!\frac{d^4 p}{(2\pi)^2}\langle
\Psi^c(p,L) \overline{\Psi}(p,L)
\rangle e^{ip(x-x')}
\frac{m}{\sqrt{\Lambda}} N(x') \right]+{\rm h.c.}, \;\;
\end{eqnarray}
where $\Psi(p,y) =  \int \! d^4x \Psi(x,y)e^{ipx}$. From this
expression, the seesaw-induced tiny mass of light Majorana neutrinos
is found in the low-energy regime ($p\to0$);
\begin{eqnarray}
M_\nu \,=\, -
\frac{m^{\rm T}}{\sqrt{\Lambda}} \,
\left. \langle i\epsilon \xi^*(p, L) \xi^\dag(p, L) \rangle \,
\frac{m}{\sqrt{\Lambda}} \right|_{p = 0}.
\label{seesaw}
\end{eqnarray}
The left-handed component $\eta(x,y)$ does not join in the seesaw
operation as it obeys the Dirichlet conditions at the
boundaries. Higher-order contributions in $p^2$ are negligible if the
energy scale of interest is much smaller than the masses of
intermediate states.

The lepton-number-violating part of the correlator is obtained by
inverting the five-dimensional Dirac operator in the presence of
bulk Majorana masses. The differential equations for the propagators
become in the mixed position-momentum space
\begin{eqnarray}
&& \hspace*{-15mm} \Big[ 
p^2 - m_d^2 - |M|^2 + \partial^2_y
- 2m_d [ \delta(y) - \delta(y - L) ] \Big]\!
\langle i\epsilon \eta^*(p, y) \eta^\dag(p, y') \rangle
= M \delta(y-y'), \;\;
\label{hpT}\\
&& \hspace*{-15mm} \Big[ 
p^2 - m_d^2 - |M|^2 + \partial^2_y
+ 2m_d [ \delta(y) - \delta(y - L) ] \Big]\!
\langle i\epsilon \xi^*(p, y) \xi^\dag(p, y') \rangle
= M \delta(y-y'). \;\;
\label{zpT}
\end{eqnarray}
Solving these equations under the boundary 
conditions $Z =1$ and $Z'=1$, we find
\begin{eqnarray}
\langle i\epsilon \eta^*(p, y) \eta^\dag(p, y') \rangle
\!\!&=&\!\! 
\frac{\sinh(qy_<) \sinh(qy_> - qL)}{q \sinh(qL)}M,
\label{propHpT} \\[.5mm]
\langle i\epsilon \xi^*(p, y) \xi^\dag(p, y') \rangle
\!\!&=&\!\! \frac{1}{(m_d^2 - q^2)q\sinh(qL)}
\Big[ q\cosh(qy_<) - m_d \sinh(qy_<) \Big] \nonumber \\[1mm]
&& \hspace*{15mm}
\times \Big[ q\cosh(qy_> - qL) - m_d \sinh(qy_> - qL) \Big]M, \qquad
\label{propZpT}
\end{eqnarray}
where $y_<$ ($y_>$) stands for the lesser (greater) of $y$ and $y'$,
and $q \equiv \sqrt{m_d^2 + |M|^2 - p^2}$. Appendix~\ref{PVBC} shows
specific details of the derivation of propagators.

With the propagators for bulk Majorana fermions at hand, the seesaw
neutrino mass is obtained by taking the low-energy limit $p\to0$ and
setting $y=y'=L$ in the propagators. Thus the neutrino mass reads
\begin{eqnarray}
M_\nu \, = \, \frac{1}{\Lambda L}
\frac{\widetilde{M}L \cosh(\widetilde{M}L)
-m_dL \sinh(\widetilde{M}L) }
{\sinh(\widetilde{M}L)}\, \frac{m^{\rm T}m}{M^*},
\label{normal}
\end{eqnarray}
that exactly agrees with the previous result (\ref{KKsum}). Once the
propagator is found, one needs neither to work on infinite-size KK
mass matrices nor to sum up the KK contributions in the seesaw formula.

\medskip

\subsection{Equivalence of two methods}
\begin{figure}[t]
\begin{center}
  \scalebox{0.52}{\includegraphics{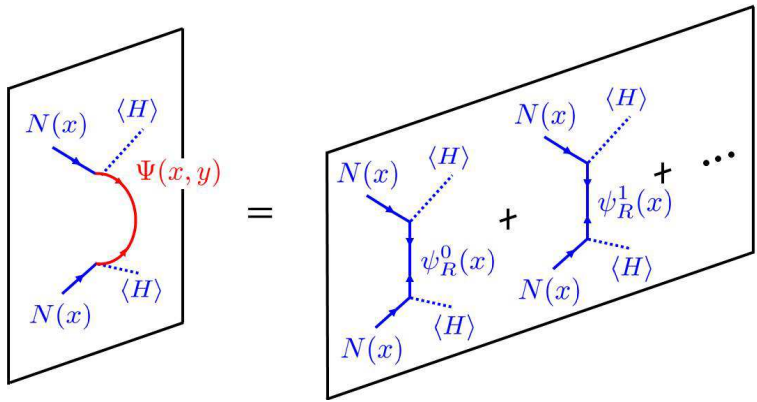}}
\end{center}
\caption{The seesaw mechanism in views of the five-dimensional
propagator and the KK summation in four-dimensional effective
theory.\medskip}
\label{fig2}
\end{figure}

The correspondence between the KK expansion and the propagator method
is clarified with the mode 
expansion $\xi(p,y)=\sum_n\chi_R^n(y)\psi_R^n(p)$:
\begin{eqnarray}
\frac{m^{\rm T}}{\sqrt{\Lambda}}\,
\langle -i\epsilon\xi^*(p,L)\xi^\dag(p,L)\rangle \,
\frac{m}{\sqrt{\Lambda}}
&=& \sum_{k,n}  
\frac{m^{\rm T}}{\sqrt{\Lambda}} \, 
\chi_R^k(L)^*
\langle -i\epsilon \psi_R^k(p)^* \psi_R^n(p)^\dag \rangle
\chi_R^n(L)^\dag \,
\frac{m}{\sqrt{\Lambda}}
\nonumber\\
&=& \sum_{k,n} m_{k}^{\rm T}
\bigg(\frac{{\cal M}_H^{\,*}}{p^2
-{\cal M}_H^{\,*}{\cal M}_H^{}}\bigg)_{\! R_{kn} }\!\! m_{n}\,,
\end{eqnarray}
where $R_{kn}$ means the matrix element for the KK right-handed
neutrinos $\psi_R^{k,n}$. Taking $p\to0$ limit exactly reproduces the
seesaw formula (\ref{seesawKK}) in the KK expansion method. A
schematic view of the correspondence is shown in Fig.~\ref{fig2}. The
propagator method avoids the jobs of mode expansion and re-summation,
and simplifies the model analysis.

The mass spectrum in the four-dimensional effective theory is found
from the poles of (\ref{propHpT}) and (\ref{propZpT}). For example,
$p^2=|M|^2$ in the denominator of (\ref{propZpT}) corresponds to the
chiral zero mode $\psi_R^0$ with the mass $M$. The other mass
eigenvalues appear at $qL=in\pi$ in (\ref{propHpT}) and
(\ref{propZpT}). These poles, $p^2 = m_d^2 + |M|^2 +
\left(\frac{n\pi}{L}\right)^2$, are consistent with the KK expansion
discussed in Section~\ref{KK}; the $n$-th $2\times 2$ block in the
heavy-sector mass matrix (\ref{KKmm}) is diagonalized as
\begin{eqnarray}
\begin{pmatrix}
-M^* & \sqrt{ m_d^2 + \left(\frac{n\pi}{L} \right)^2} \\
 \sqrt{ m_d^2 + \left(\frac{n\pi}{L} \right)^2}& M \\
\end{pmatrix}
\to 
\begin{pmatrix}
 \sqrt{ m_d^2 + |M|^2 + \left(\frac{n\pi}{L} \right)^2} & \\
 & -\sqrt{ m_d^2 + |M|^2 + \left(\frac{n\pi}{L} \right)^2} \\
\end{pmatrix}.\nonumber
\end{eqnarray}
Thus the propagator method also reduces the efforts of expansion and
(generally complicated) diagonalization to have physical mass 
eigenvalues.

\medskip

\section{Applications of the propagator}
\label{appli}
In this section, we apply the propagator to the higher-dimensional 
seesaw model described in Section~\ref{base}. We discuss neutrino-mass
phenomenology under various types of boundary conditions, applications
to flavor models, and the seesaw mechanism in the warped geometry.

\subsection{The standard seesaw-induced mass}
For the standard boundary condition ($Z=1$ and $Z'=1$), the neutrino
mass after the seesaw operation is given by (\ref{normal}). Let us
first study a simple case of vanishing bulk Dirac mass $m_d =0$. The
seesaw-induced mass becomes
\begin{eqnarray}
M_\nu \, = \,
\frac{1}{\Lambda L}\, \frac{|M|L}{\tanh\left( |M|L \right)}
\frac{m^{\rm T}m}{M^*}.
\label{normal2}
\end{eqnarray}
The effect of extra dimension is evident in the appearance of
hyperbolic factor. If the contribution from the KK-excited modes is 
negligible, the seesaw with only the zero mode gives a usual
four-dimension-like formula
\begin{eqnarray}
M_\nu^{(0)} \,=\, \frac{m_0^{\rm T }m_0}{M^*} \,=\, 
\frac{1}{\Lambda L}\frac{m^{\rm T}m}{M^*}.
\label{0seesaw}
\end{eqnarray}
To see how the zero-mode approximation is related to the
complete formula (\ref{normal2}), consider two extreme cases
in (\ref{normal2}); 
\begin{eqnarray}
\begin{array}{ll}
M_\nu  \,\simeq\,  \frac{1}{\Lambda L}\frac{m^{\rm T}m}{M^*}
& \quad {\rm for} \quad ML \ll 1, \\[2mm]
M_\nu  \,\simeq\,  \frac{1}{\Lambda L}\frac{m^{\rm T}m}{1/L}
& \quad {\rm for} \quad ML \gg 1.
\end{array}
\label{approx}
\end{eqnarray}
In the former limit, the KK-excited modes are decoupled, more exactly,
each nonzero KK level has much smaller Majorana mass (lepton number
violation) than its mass eigenvalue and gives negligible 
contribution to the seesaw Majorana mass. The result hence
coincides with the zero-mode seesaw (\ref{0seesaw}). In the latter
case, the effects of higher KK modes ($n'\gtrsim ML$) are
dropped due to the same reason as the former limit, and the lower-mode
contributions are piled up to 
giving $\frac{1}{M}\times n' \simeq \frac{1}{1/L}$. In any case, the
heavy mass scale in the seesaw mechanism is determined by a smaller
one between $M$ and $1/L$. If the neutrino Yukawa couplings are of
order unity, $\Lambda > 10^{14}\,{\rm GeV}$ 
($\Lambda \sim 10^{14}\,{\rm GeV}$) for ${\cal O}({\rm eV})$ neutrino
masses in the former (latter) case. The other two 
scales, $M$ and $1/L$, vary widely depending on the cutoff $\Lambda$.

For the full expression (\ref{normal}) including the bulk Dirac
mass, an interesting case is that $m_d$ is much larger than the
other mass scales in the bulk, $M$ and $1/L$. In the 
regime $m_d\gg M$ and $\widetilde{M}\gg1/L$, the seesaw-induced mass
turns out to be
\begin{eqnarray}
M_\nu \,\simeq\, \frac{1}{\Lambda L } \,\frac{M}{m_d}\,
\frac{m^{\rm T}m}{1/L}.
\label{en1}
\end{eqnarray}
The obtained neutrino mass decreases as the lepton number violation
(the right-handed Majorana neutrino mass) decreases. That is the
opposite behavior to the usual seesaw mechanism. In view of the KK
expansion, the inverse seesaw~\cite{inv} takes place in each KK level.

An essential point for realizing the inverse seesaw is pairing
two KK neutrinos in each level into one pseudo Dirac fermion. That is,
Majorana masses of two spinors should be much smaller than their 
lepton-number-conserving Dirac mixing. It seems unlikely that the
inverse seesaw occurs since a chiral zero mode exists and does not
belong to the vector-like KK tower. However the zero mode has a
localized wavefunction which can make its seesaw contribution
irrelevant. In the regime $m_d\gg M$ and $\widetilde{M}\gg 1/L$, the
bulk Dirac mass $m_d$ is large so that the zero mode is localized
towards the $y=0$ boundary and its bulk--boundary interaction 
at $y=L$ is exponentially suppressed. On the other hand, the
KK-excited modes have sizable bulk--boundary interactions compared to
the zero mode [see the wavefunctions (\ref{KKmm2})]. Subtracting the
zero mode, the KK mass matrix $\mathcal{M}$ is effectively written as
\begin{eqnarray}
\mathcal{M} \,\simeq\,
\left( \begin{array}{c|ccc}
 & m_1^{\rm T} &  & \cdots \\\hline
m_1 & -M^* & M_{K_{1}} &  \\
 &  \phantom{-}M_{K_{1}} & M &  \\
\vdots &  &  & \ddots \\
\end{array}
\right).
\label{inverseSS}
\end{eqnarray}
Mode by mode, the inverse seesaw takes place if $M_{K_n} = 
\sqrt{m_d^2 + (n\pi/L)^2} \gg |M|$. The inverse seesaw effect is not
available unless the zero mode is localized away from the SM
boundary. This is consistent with the fact that a positive $m_d$ is
needed for (\ref{en1}) in the propagator method.
 
For the inverse seesaw case (\ref{en1}), a large scale such as the
Planck or grand unification scale is not necessary for producing
eV-scale neutrino masses. For example, if $\Lambda$ and $m_d$ are
around TeV, the bulk Majorana mass is $M \sim 10^2 \,{\rm eV}$ for the
neutrino Yukawa coupling of order unity. Thus the model could be 
reconciled with low-cutoff scenarios such as the large extra 
dimensions.\footnote{In this case, the present setup is
interpreted as a subspace of higher dimensions not to conflict with
experimental bounds.}

\medskip

\subsection{Twisted boundary conditions}
Next we study other types of boundary conditions for bulk
right-handed neutrinos. The seesaw setup is the same as before
(Fig.~\ref{fig1}). Let us consider the case $Z=-1$ and $Z'=+1$,
namely, different reflection profiles are assigned at two
boundaries. In this case, a non-trivial 
twist {\it \`a la} Scherk-Schwarz~\cite{SS} is generated by the
translation along the extra dimension; $T=Z'Z=-1$. The seesaw mass
formula (\ref{seesaw}) is unchanged and only difference is the form of
propagator. Using the propagator presented in Appendix~\ref{PVBC},
we find the seesaw-induced neutrino mass for $Z=-1$ and $Z'=+1$,
\begin{eqnarray}
M_\nu \,=\,
\frac{1}{\Lambda L}\,
\frac{(|M|L)^2 \sinh(\widetilde{M}L)}{\widetilde{M}L\cosh(\widetilde{M}L) 
+ m_dL \sinh(\widetilde{M}L) }
\frac{m^{\rm T}m}{M^*}.
\end{eqnarray}
A vanishing Dirac mass, $m_d=0$, reveals an essential difference from
the standard boundary condition. For the present twisted boundary
condition, we have
\begin{eqnarray}
M_\nu \,=\, \frac{1}{\Lambda L}\,
|M|L\tanh\left( |M|L \right)
\frac{m^{\rm T}m}{M^*}.
\label{ap}
\end{eqnarray}
Contrary to (\ref{normal2}), the neutrino mass has a tanh factor in
the numerator and behaves as
\begin{eqnarray}
\begin{array}{ll}
M_\nu \,\simeq\, \frac{1}{\Lambda L}(ML)^2 \frac{m^{\rm T}m}{M}
& \quad {\rm for} \quad ML \ll 1, \\[2mm]
M_\nu \,\simeq\, \frac{1}{\Lambda L}\frac{m^{\rm T}m}{1/L}
& \quad {\rm for} \quad ML \gg 1. \\
\end{array}
\label{approx2}
\end{eqnarray}
The latter case [$\tanh(|M|L)\simeq 1$] leads to the same result as
the standard one. This is because, in the large-size limit of extra
dimension, the difference of boundary conditions at $y=0$ is
irrelevant to the physics at $y=L$ where the SM fields reside. The
former case $ML \ll 1$ shows up an interesting feature of the twisted
boundary condition; the seesaw-induced mass is proportional to 
the Majorana mass parameter of bulk heavy neutrinos.

In terms of the KK expansion, such unusual seesaw behavior is
understood as the inverse seesaw suppression, which is similar to the
previous standard case (\ref{en1}). In the present case, the
inverse seesaw is achieved by the twisted boundary condition which
forbids no-winding wavefunctions. That is, the zero mode is absent in
the effective theory. (In the standard case, the zero-mode effect is
suppressed by its localized wavefunction.) With the twisted boundary 
condition $Z = -1$ and $Z' = +1$, the KK wavefunctions and masses are
given by
\begin{eqnarray}
\chi_R^n \,=\, \sqrt{\frac{2}{L}} 
\cos\left( M_{K_n} y \right),\qquad
\chi_L^n \,=\, \sqrt{\frac{2}{L}}
\sin\left( M_{K_n} y \right),\qquad
M_{K_n} \,=\, \left( n-\frac{1}{2} \right)\frac{\pi}{L},
\label{sincos1/2}
\end{eqnarray}
for $n \geq 1$. The neutrino mass matrix is then given by the same
form as (\ref{inverseSS}) and the inverse seesaw takes place 
if $ML \ll 1$.

Table 1 shows the seesaw-induced neutrino mass $M_\nu$ for various
limits and boundary conditions. The columns represent two patterns of
boundary conditions $(Z,Z')$ and the rows possible hierarchies among
the mass parameters $M$, $1/L$, and $m_d$. The other two 
conditions $(Z,Z') = (+1,-1)$ and $(-1,-1)$ correspond to the
exchanges of right- and left-handed components of bulk 
fermions. If $1/L$ is the largest, it turns out that the neutrino
mass does not depend on the ordering of $M$ and $m_d$. Further, 
if $1/L$ is not the largest, only the hierarchy 
between $M$ and $m_d$ affects the results. Therefore three ordering
patterns (labeled by Type~A,~B,~C in Table~\ref{pattern}) are
practically relevant. The entries with symbols~``$\,\star\,$'' are
viable patterns for the inverse seesaw suppression.

Out of six general possibilities, three patterns have suppression
factors by the inverse seesaw. Two of them are already discussed:
Type~A with the standard condition [Eq.~(\ref{en1})] and Type~B 
with the opposite parity at the distant brane
[Eq.~(\ref{approx2})]. The third case belongs to the Type~A hierarchy
with the twisted boundary condition, and results in the same form as
the non-twisted case. In any case, a small bulk Majorana mass $M$ is a
key to obtain suppressed neutrino masses.

\begin{table}
\begin{center}
\begin{tabular}{l|c|c}\hline\hline
& $(Z,Z')=(+1,+1)$ & $(Z,Z')=(-1,+1)$ \\\hline
\renewcommand{\arraystretch}{2.0}
\begin{tabular}{c}
Type A:~~~$m_d \,\gg\, M, 1/L$ \\
\end{tabular}
&  $\star$ $\frac{1}{\Lambda L} \frac{M}{m_d} \frac{m^2}{1/L}$  
&  $\star$ $\frac{1}{\Lambda L} \frac{M}{m_d} \frac{m^2}{1/L}$  
\\ \hline
\renewcommand{\arraystretch}{2.0}
\begin{tabular}{c}
Type B:~~~$1/L \,\gg\, M, m_d$ \\
\end{tabular}
&  $\frac{1}{\Lambda L}\frac{m^2}{M}$
&  $\star$ $\frac{1}{\Lambda L} (ML)^2 \frac{m^2}{M}$ 
\\ \hline
\renewcommand{\arraystretch}{2.0}
\begin{tabular}{c}
Type C:~~~$M \,\gg\, m_d, 1/L$ \\
\end{tabular}
&  $\frac{1}{\Lambda L}\frac{m^2}{1/L}$  
&  $\frac{1}{\Lambda L}\frac{m^2}{1/L}$   
\\ \hline
\end{tabular} 
\end{center}
\caption{The seesaw-induced neutrino masses $M_\nu$ in various
situations. The entries with symbols~``$\,\star\,$'' have further
suppression factors beyond the standard seesaw up to the volume
factor $1/\Lambda L$.\medskip}
\label{pattern}
\end{table}

\medskip

\subsection{Flavor symmetry breaking}
\label{tbf}
We have so far focused on the eigenvalues of seesaw-induced masses. The 
higher-dimensional seesaw also gives an interesting possibility for the
generation structure of light neutrinos, e.g., the boundary condition
breaking of flavor symmetry~\cite{TF}. We will show in this section
that the propagator method simplifies the previous KK-mode analysis
and also makes it clear how flavor symmetry is broken down.

The setup is the same as before and the Lagrangian is written down 
with the generation indices ($i,j = 1,2,3$)
\begin{eqnarray}
\mathcal{L}_{\rm bulk} &=&
i\overline{\Psi}_j \Gamma^M \partial_M \Psi_j
- m_{d_{ij}}\theta(y)\overline{\Psi}_i  \Psi_j 
- \frac{1}{2} (M_{ij} \overline{\Psi^c}_i\Psi_j + {\rm h.c.}), \\
\mathcal{L}_{\rm boundary} &=&
-\left(\, \frac{m_{ij}}{\sqrt{\Lambda}} \overline{\Psi}_iN_j  
\,+\, {\rm h.c.} \right) \delta(y - L).
\end{eqnarray}
The general boundary conditions are
\begin{eqnarray}
\Psi_i(x,-y) = (Z_{ij} \otimes \gamma_5) \Psi_j(x,y), \qquad
\Psi_i(x,L- y) = (Z'_{ij} \otimes \gamma_5) \Psi_j(x,L+y). 
\end{eqnarray}
When the Lagrangian is invariant under some flavor 
symmetry, $Z$ and $Z'$ are allowed to be identified as some elements
of the symmetry group. The matrices $Z$ and $Z'$ represent the parity
operations in the field space and should satisfy $Z^2=I$ and $Z'^2=I$. 

As an example of flavor symmetry, we adopt the $S_3$ permutation,
which has been widely studied in the 
literature~\cite{S3}. The $S_3$ group is the simplest non-abelian
discrete group with six elements: the identity $I$, two cyclic
permutations $R_{1,2}$, and three permutations $P_{1,2,3}$. The
irreducible representations are the doublet $\underline{2}$, 
pseudo singlet $\underline{1}'$ and singlet $\underline{1}$. The
representation matrices 
for $\underline{3}=\underline{2} + \underline{1}$ are given by
\begin{eqnarray}
&&
I \;\,=\begin{pmatrix} 1 & & \\ & 1 & \\ & & 1 \end{pmatrix},\quad
R_1 =\begin{pmatrix} & & 1 \\ 1 & & \\ & 1 & \end{pmatrix},\quad
R_2 =\begin{pmatrix} & 1 & \\ & & 1 \\ 1 & & \end{pmatrix}, 
\nonumber \\[1mm]
&&
P_1 =\begin{pmatrix} 1 & & \\ & & 1 \\ & 1 & \end{pmatrix},\quad
P_2 =\begin{pmatrix} & & 1 \\ & 1 & \\ 1 & & \end{pmatrix},\quad
P_3 =\begin{pmatrix} & 1 & \\ 1 & & \\ & & 1 \end{pmatrix}.
\end{eqnarray}
The three generations are treated democratically under the flavor
symmetry, and its breaking is a key to account for the observed mass
differences and mixing angles in the lepton sector.

Suppose that the bulk fields $\Psi_i(x,y)$ and the SM 
neutrinos $N_i(x)$ belong to $\underline{3}$ representations 
of $S_3$. The symmetry-invariant mass parameters are given by 
the combinations of the identity $I$ and the democratic 
matrix $D$ whose all elements are one thirds:
\begin{eqnarray}
M \,=\, M_1 I + M_2 D, \qquad
m_d \,=\, \delta_1 I + \delta_2 D, \qquad
m \,=\, \mu_1 I + \mu_2 D.
\label{baremass}
\end{eqnarray}
Each mass matrix is described by two parameters, which reflects the
fact that the tensor product of two $\underline{3}$'s contains two
singlet components.

Now let us fix the boundary condition by identifying $Z$ and $Z'$ as
the $S_3$ group elements. Notice that the cyclic 
permutations $R_{1,2}$ do not satisfy the parity conditions and are
excluded. We then consider
\begin{eqnarray}
Z \,=\, P_1, \qquad Z' \,=\, I.
\label{zzp4}
\end{eqnarray}
The translation is twisted as $T=Z'Z=P_1$ and becomes a typical
(discrete) example of the original Scherk-Schwarz
theory~\cite{SS}. The boundary conditions (\ref{zzp4}) imply that the
flavor symmetry breaking occurs in a separate place from the
(symmetry-preserving) SM boundary and is mediated by bulk right-handed
neutrinos through the seesaw mechanism.

It is convenient to move onto the basis where $P_1$ and the mass
matrices (\ref{baremass}) are diagonal; they 
become $M' = {\rm diag}(M_1,M_1,M_1+M_2)$, 
$m_d' = {\rm diag}(\delta_1,\delta_1,\delta_1+\delta_2)$, 
$m' = {\rm diag}(\mu_1,\mu_1,\mu_1+\mu_2)$, and 
the boundary conditions are given 
by $Z={\rm diag}(1,-1,1)$ and $Z'={\rm diag}(1,1,1)$. It is
straightforward in this basis to perform the five-dimensional seesaw
using the results in Section~\ref{prop} and Appendix~\ref{PVBC}\@. On
the other hand, the propagator in the original basis is useful for 
intuitive understanding of the boundary condition breaking of flavor
symmetry. It turns out that
\begin{eqnarray}
\langle i\epsilon \xi^*(p, y) \xi^\dag(p, y') \rangle
&=& 
Z_p^{++}(y,y',\delta_1+\delta_2,M_1+M_2)\, D \nonumber\\
&& \qquad\;\; - Z_p^{++}(y,y',\delta_1,M_1)\, E 
 \,+\, Z_p^{-+}(y,y',\delta_1,M_1)\, F, \quad
\label{proptwist}
\end{eqnarray}
where 
\begin{eqnarray}
E \,\equiv\, \frac{1}{6}\begin{pmatrix}
4 &-2&-2 \\
-2&1 & 1 \\
-2&1 & 1 \\
\end{pmatrix},
\qquad
F \,\equiv\, \frac{1}{2}\begin{pmatrix}
&& \\
&1&-1 \\
~&-1&1 \\
\end{pmatrix}.
\end{eqnarray}
Here the notation for $Z_p$'s follows Appendix~\ref{PVBC}\@. Flavor
symmetry breaking is clearly seen in (\ref{proptwist}); the 
matrices $E$ and $F$ are not invariant under the general permutations,
while they are invariant under the exchange of the second and third
generations. With the twisted boundary condition imposed, the 
original $S_3$ is broken down to $S_2$.

The seesaw-induced mass $M_\nu$ is computed by taking the low-energy 
limit $p\to0$ and setting $y= y' = L$ in the propagator
(\ref{proptwist}), and by multiplying the boundary mass 
matrix $m$ given in (\ref{baremass}). For a simple case with vanishing
bulk Dirac mass, $M_\nu$ reads
\begin{eqnarray}
M_\nu \,=\, 
\frac{\frac{M_1+M_2}{|M_1+M_2|}(\mu_1+\mu_2)^2}
{\Lambda\tanh(|M_1+M_2|L)}D
+ 
\frac{\frac{M_1}{|M_1|}(\mu_1)^2}{\Lambda\tanh(|M_1|L)}E
+
\frac{\frac{M_1}{|M_1|}(\mu_1)^2}{\Lambda\coth(|M_1|L)}F. \;
\label{maj}
\end{eqnarray}
The mass matrix has the same structure of generations mixing as the
propagator. An important property of (\ref{maj}) is that it is
diagonalized by the tri-bimaximal mixing matrix~\cite{TBM}
\begin{eqnarray}
V_{\rm tri} \,=\, \begin{pmatrix}
\frac{-2}{\sqrt{6}} & \frac{1}{\sqrt{3}} & 0 \\
\frac{1}{\sqrt{6}} & \frac{1}{\sqrt{3}} & \frac{-1}{\sqrt{2}} \\
\frac{1}{\sqrt{6}} & \frac{1}{\sqrt{3}} & \frac{1}{\sqrt{2}} \\
\end{pmatrix},
\end{eqnarray}
for which various theoretical approaches have been
discussed~\cite{TBMmodel}. Notice that the diagonalization is free from
the parameters involved in the Lagrangian, and the tri-bimaximal
mixing is a rigid prediction of the flavor twisting. The prediction is
not disturbed by nonzero bulk Dirac masses since the flavor structure
is independent of the parameters $\delta_{1,2}$ [see the propagator
(\ref{proptwist})].

The boundary condition (\ref{zzp4}) is an example of all possible
choices. However the tri-bimaximal mixing is also induced by many
other types of boundary conditions~\cite{TF}. Thus the prediction is
not a special feature of (\ref{zzp4}), but it is rather common outcome
of the present setup and twisted flavors.

\medskip

\subsection{Seesaw in the warped geometry}
The higher-dimensional seesaw is calculable with the propagator method
not only in the flat space but also for a generic class of curved
geometry. Even if the background is so complicated that KK
wavefunctions and mass eigenvalues cannot be found, the seesaw-induced
neutrino mass is analytically obtained.

Let the five-dimensional seesaw setup be placed on the gravitational
background
\begin{eqnarray}
ds^2 \,=\, e^{-2k|y|} \eta_{\mu\nu} dx^\mu dx^\nu - dy^2,
\end{eqnarray}
where $k$ stands for the anti de-Sitter curvature 
and $\eta_{\mu\nu}$ is the four-dimensional Minkowski metric. The
previous Lagrangian is modified such that it incorporates the gravity;
\begin{eqnarray}
\mathcal{L} \;=\, \sqrt{g} \bigg[
i\overline\Psi \cancel{D} \Psi
-  m_d \theta(y) \overline\Psi\Psi
- \Big(\, \frac{1}{2} M\overline{\Psi^c} \Psi 
+\frac{m}{\sqrt{\Lambda}} \overline{\Psi} N\delta(y - L) 
+ {\rm h.c.} \Big) \bigg].
\end{eqnarray}
The covariant derivative $\cancel{D}$ includes the spin connection,
and the generation indices are suppressed. The SM fields are assumed
to reside in the $y=L$ boundary to solve the gauge hierarchy problem
with the warp factor~\cite{RS}. In the above Lagrangian, the SM
neutrinos $N$ and the Higgs field $H$ have been rescaled for their
kinetic terms being canonical, and therefore $m$ is a parameter of the 
electroweak scale.

The non-trivial metric factor modifies the equations of
propagators. After the (non-canonical) 
rescaling $\Psi \to e^{2k|y|}\Psi$, the lepton-number-violating parts
of bulk neutrino propagators are determined by
\begin{eqnarray}
&& \Big[
e^{2k|y|}p^2 - m_d^2 - MM^* + \partial^2_y
- 2m_d [ \delta(y) - \delta(y - L) ] \Big]
\langle i\epsilon \eta^*(p, y) \eta^\dag(p, y') \rangle \nonumber\\
&& \hspace*{55mm}
- k \theta(y) e^{k|y|} p^\mu \sigma_\mu
\langle i\epsilon \xi^*(p, y) \xi^\dag(p, y') \rangle
\,=\, M \delta(y - y'), \qquad\quad
\label{hpW}\\[1mm]
&& \Big[ 
e^{2k|y|}p^2 - m_d^2 - MM^* + \partial^2_y
+ 2m_d [ \delta(y) - \delta(y - L) ] \Big]
\langle i\epsilon \xi^*(p, y) \xi^\dag(p, y') \rangle \nonumber\\
&& \hspace*{55mm}
+ k \theta(y) e^{k|y|}  p^\mu \bar{\sigma}_\mu
\langle i\epsilon \eta^*(p, y) \eta^\dag(p, y') \rangle
\,=\, M \delta(y - y').
\label{zpW}
\end{eqnarray}
Unlike (\ref{hpT}) and (\ref{zpT}) in the flat background, these are
the coupled equations due to the non-vanishing curvature. Further, in
the presence of the exponential factor, it seems difficult to solve
the above equations. However the low-energy behavior ($p\to0$) of the
solutions is sufficient for the seesaw mechanism.\footnote{To be
precise, the following procedure is valid if, at any point in the
bulk, $e^{k|y|}p$ is smaller than the fundamental scale of the theory.} 
It is found from (\ref{hpW}) and (\ref{zpW}) that, in the low-energy
limit, the warp factor vanishes away from the problem and the
propagators (with the non-canonical rescaling) are found to have the
same forms as in the flat extra dimension. Notice that the low-energy
limit $p\to0$ is allowed before solving the propagator equations only
if the solutions are non-singular in that limit. The regularity is
ensured in the seesaw theory where the bulk Majorana mass lifts the
chiral zero modes (right-handed neutrinos) which are otherwise
massless even in the presence of bulk Dirac masses. In the end, the
seesaw-induced mass in the warped geometry is evaluated with the
propagator in the flat space and the couplings in the rescaled basis.

The procedure for acquiring $M_\nu$ goes parallel to the flat
case. The only difference is the appearance of warped metric factors,
which count the mass dimensions of couplings. Let us incorporate the
generation structure as before by supposing that the Lagrangian
respects the $S_3$ flavor symmetry and the bulk fermions obey the
boundary conditions $Z=P_1$ and $Z'=I$. The flavor symmetry requires
the Lagrangian mass parameters $M$, $m_d$, and $m$ to have the form
(\ref{baremass}). After all, the seesaw-induced mass matrix is found
(for vanishing bulk Dirac masses)
\begin{eqnarray}
M_\nu \,=\, 
\frac{\frac{M_1+M_2}{|M_1+M_2|}(\mu_1+\mu_2)^2}
{\Lambda'\tanh(|M_1+M_2|L)}D
+ 
\frac{\frac{M_1}{|M_1|}(\mu_1)^2}{\Lambda'\tanh(|M_1|L)}E
+
\frac{\frac{M_1}{|M_1|}(\mu_1)^2}{\Lambda'\coth(|M_1|L)}F, \;
\end{eqnarray}
where $\Lambda' = \Lambda e^{-kL}$. Comparing this to the previous
result (\ref{maj}), one finds that the warped geometry modifies the
neutrino mass only by an overall factor of each matrix, and the
details of propagators (or KK wavefunctions) do not affect the flavor
structure of low-energy neutrinos. That is the geometry-free nature of
seesaw-induced masses in higher dimensions~\cite{WY}. The conclusion
is unchanged by non-vanishing bulk Dirac masses.

If the warp factor is used to solve the gauge hierarchy problem, the
effective seesaw scale $\Lambda'$ is around TeV and the neutrino mass
of ${\cal O}({\rm eV})$ requires tiny values of neutrino Yukawa
couplings. A way to ameliorate this problem is to consider bulk
Majorana masses of intermediate scale which generate additional
suppression via the inverse seesaw. For example, 
when $M,\,1/L \ll m_d$ (Type A in Table~\ref{pattern}), the neutrino
masses are given 
by $m_{1,3} \sim \frac{M_1}{\delta_1}\frac{\mu_1^2}{\Lambda'}$ and 
$m_2 \sim \frac{M_1+M_2}{\delta_1+\delta_2}
\frac{(\mu_1+\mu_2)^2}{\Lambda'}$. Thus a small 
ratio $M/\delta$ is used to make a tuning of Yukawa couplings
reduced. Another way to have mild Yukawa hierarchy is to extent the
SM neutrinos (the lepton doublets) into the extra dimension and to
utilize the localization effect. It is however noted that the
wavefunction suppression by the left-handed neutrinos cannot be
arbitrarily strong as it also brings down the charged-lepton mass
scale. For example, in the case that the right-handed tau resides on
the SM boundary, its wavefunction lowers the neutrino masses by the
factor of $(m_\tau/\Lambda')^2$.

\medskip

\section{Summary}
We have studied the higher-dimensional seesaw mechanism with two
methods: the KK-mode expansion and the five-dimensional
propagators. The propagator is derived for various types of
boundary conditions and mass parameters of bulk right-handed
neutrinos. The propagator method simplifies the calculation of seesaw
induced masses and clarifies the physical implications. That can skip
identifying KK eigenfunctions, evaluating (infinite-dimensional) mass
matrices, and summing up the KK contributions to the seesaw-induced
mass. Noticing that the neutrino mass is estimated in the low-energy
limit, its explicit form is obtained even when the background geometry
is non-trivial and a suitable KK expansion is not viable. The
propagator method is also useful to capture symmetry-breaking effects
by boundary conditions. As an application of these facts, we have
discussed the Scherk-Schwarz breaking of flavor symmetry in the flat
and warped extra dimensions. The neutrino mass matrix in the warped
case is calculated in the same fashion as in the flat case, with the
same propagator and rescaled couplings. The two results differ
only by the overall metric factor.

The higher-dimensional seesaw realizes various structures 
in low-energy effective theory, in particular, suitable for the
inverse seesaw suppression of neutrino masses. For instance, by taking
the bulk Dirac mass such that the zero-mode wavefunction is localized
away from the SM fields, its seesaw contribution is suppressed and the
seesaw mediator is played by vector-like pairs of KK-excited modes
with almost Dirac nature. In this case, the seesaw-induced mass is
proportional to the (lepton-number-violating) bulk Majorana
mass. Alternatively, the Dirichlet boundary condition for right-handed
neutrinos forbids the existence of zero mode and the inverse seesaw is
realized naturally. The possible forms of seesaw-induced mass in
various limits are summarized in Table~\ref{pattern}.

Besides several examples discussed in this paper, there may be other
broad usages of the (lepton-number-violating) propagator in
higher-dimensional theory, e.g., for seesaw collider
phenomenology~\cite{5d_nu_collider}, leptogenesis~\cite{leptogen}, and
so on. Such phenomenological applications remains to be studied in
future work.

\medskip

\subsection*{Acknowledgments}
A.W. would like to thank Toshifumi Yamashita for helpful discussion.
The authors are supported in part by the scientific grant from the
ministry of education, science, sports, and culture of Japan
(No.~20740135, No.~21340055) and also by the grant-in-aid for the global COE program
"The next generation of physics, spun from universality and emergence".
\bigskip

\appendix

\section{Lorentz spinors and gamma matrices}
\label{LG}
In this work, the gamma matrices are taken as
\begin{gather}
\{ \Gamma^M, \Gamma^N \} \,=\, 
2\eta^{MN} \,=\, 2\,{\rm diag}(+1, -1,-1,-1,-1), \\[2mm]
\Gamma_\mu = \gamma_\mu = 
\begin{pmatrix}
 & \sigma_\mu \\
 \bar{\sigma}_\mu & \\
\end{pmatrix},\qquad
i\Gamma_4 = \gamma_5 =  
\begin{pmatrix}
1 &  \\
  & -1 \\
\end{pmatrix},
\end{gather}
where $\sigma_\mu=(1,\sigma_i)$ and $\bar{\sigma}_\mu=(1,-\sigma_i)$.
A 4-component spinor is written in terms of 2-component spinors as
\begin{eqnarray}
\Psi \,=\, \begin{pmatrix}
\xi_\alpha \\
\eta^{\dot{\alpha}} \\
\end{pmatrix}.
\end{eqnarray}
The Dirac and charge conjugates for $\Psi$ are given by 
\begin{eqnarray}
\overline{\Psi} \,=\, 
\big(\eta^{*\alpha} \;\, \xi_{\dot{\alpha}}^* \big), \qquad\quad
\Psi^c \,=\, C_5 \overline{\Psi}^{\rm T} \,=\, \begin{pmatrix}
-\epsilon_{\alpha\beta}\eta^{*\beta} \\[.5mm]
-\epsilon^{\dot{\alpha}\dot{\beta}}\xi^*_{\dot{\beta}}\,
\end{pmatrix},
\end{eqnarray}
where $C_5$ is the charge conjugation matrix in five 
dimensions: $C_5=i\gamma^2\gamma^0\gamma_5$.
The antisymmetric tensors are
\begin{eqnarray}
\epsilon^{\alpha\beta} = \epsilon_{\alpha\beta} = 
\epsilon^{\dot{\alpha}\dot{\beta}} = \epsilon_{\dot{\alpha}\dot{\beta}}
 = \begin{pmatrix} & 1 \\ -1 &  \\ \end{pmatrix}.
\end{eqnarray}

\medskip

\section{Propagators for bulk Majorana fermions}
\label{PVBC}
To find the lepton-number-violating part of the propagator, it is
convenient to rewrite the bulk Lagrangian (\ref{Lbulk}) as
\begin{eqnarray}
\mathcal{L}_{\rm bulk} \,=\, \frac{1}{2} 
\left(
\overline{\Psi}\,\,\,\overline{\Psi^c} \right) \hat{D}
\begin{pmatrix}
\Psi \\
\Psi^c \\
\end{pmatrix} ,
\end{eqnarray}
\begin{eqnarray}
\hat{D} \,=\, 
\begin{pmatrix}
i\cancel{\partial} - \gamma_5 \partial_y - m_d\theta(y) & -M^* \\
-M & i\cancel{\partial} - \gamma_5 \partial_y + m_d\theta(y)
\end{pmatrix}.
\end{eqnarray}
The propagator is given by the inverse of $\hat{D}$;
\begin{eqnarray}
\hat{D} G(x,x',y,y') \,=\, i\delta^4(x-x')\delta(y-y'),
\label{inverse}
\end{eqnarray}
where
\begin{eqnarray}
G(x,x',y,y') &=& \begin{pmatrix}
\langle \Psi(x,y)\overline{\Psi}(x',y') \rangle & 
\langle \Psi(x,y)\overline{\Psi^c}(x',y') \rangle \\[1mm]
\langle \Psi^c(x,y)\overline{\Psi}(x',y') \rangle & 
\langle \Psi^c(x,y)\overline{\Psi^c}(x',y') \rangle
\end{pmatrix} \\[2mm]
&=& 
\left(\begin{array}{cc|cc}
\langle \xi \eta^\dag \rangle &
\langle \xi \xi^\dag \rangle & 
\langle \xi \xi^{\rm T}\epsilon \rangle &
\langle \xi \eta^{\rm T}\epsilon \rangle \\
\langle \eta \eta^\dag \rangle & 
\langle \eta \xi^\dag \rangle &
\langle \eta \xi^{\rm T}\epsilon \rangle &
\langle \eta \eta^{\rm T}\epsilon \rangle \\ \hline
\langle -\epsilon \eta^* \eta^\dag \rangle &
\langle -\epsilon \eta^* \xi^\dag \rangle &
\langle -\epsilon \eta^* \xi^{\rm T} \epsilon \rangle &
\langle -\epsilon \eta^* \eta^{\rm T} \epsilon \rangle \\
\langle -\epsilon \xi^* \eta^\dag \rangle &
\langle -\epsilon \xi^* \xi^\dag \rangle &
\langle -\epsilon \xi^* \xi^{\rm T} \epsilon \rangle &
\langle -\epsilon \xi^* \eta^{\rm T} \epsilon \rangle
\end{array}\right).
\end{eqnarray}
The upper-right and the lower-left blocks violate the lepton
number. These two blocks are related 
as $\langle \Psi(x,y)\overline{\Psi^c}(x',y') \rangle =
\Gamma^0 \langle \Psi^c(x,y)\overline{\Psi}(x',y') \rangle^\dag 
\Gamma^0 \big|_{x,y \leftrightarrow x',y'}$.

The equation (\ref{inverse}) is written in the mixed position-momentum
space as
\begin{eqnarray}
\big(\cancel{p} - \gamma_5\partial_y - m_d\theta(y) \big)
\langle \Psi(p,y)\overline{\Psi}(p,y') \rangle
- M^* \langle \Psi^c(p,y)\overline{\Psi}(p,y') \rangle
&=& i\delta(y-y'), \quad  \\
\big(\cancel{p} - \gamma_5\partial_y + m_d\theta(y) \big)
\langle \Psi^c(p,y)\overline{\Psi}(p,y') \rangle
- M \langle \Psi(p,y)\overline{\Psi}(p,y') \rangle
&=& 0.
\end{eqnarray}
By eliminating the lepton-number-conserving 
part $\langle\Psi\overline{\Psi}\rangle$, one obtains
\begin{eqnarray}
\big[ p^2 - m_d^2 - MM^* + \partial^2_y
+ 2m_d [ \delta(y) - \delta(y - L) ] \big]
Z_p(y,y') \,&=& M \delta(y - y'),
\label{zp} \\[.5mm]
\big[ p^2 - m_d^2 - MM^* + \partial^2_y
- 2m_d [ \delta(y) - \delta(y - L) ] \big]
H_p(y,y') &=& M \delta(y - y').  \qquad
\label{hp}
\end{eqnarray}
Here we have introduced the notation
\begin{eqnarray}
Z_p(y,y') \,\equiv\, \langle i\epsilon\xi^*(p,y)\xi^\dag(p,y')\rangle,
\qquad
H_p(y,y') \,\equiv\, \langle i\epsilon\eta^*(p,y)\eta^\dag(p,y')\rangle.
\end{eqnarray}
The general solutions in the bulk are 
\begin{eqnarray}
Z_p(y,y') &=& A_Z(y')\sinh(qy)+B_Z(y')\cosh(qy),\\
H_p(y,y') &=& A_H(y')\sinh(qy)+B_H(y')\cosh(qy),
\end{eqnarray}
with $q=\sqrt{m_d^2 + MM^* - p^2}$. The 
coefficients $A_{Z,H}$ and $B_{Z,H}$ are determined by the boundary
conditions and matching in the following.

Let us first consider $Z = +1$ and $Z' = +1$ that the
right-(left-)handed component obeys the Neumann (Dirichlet) conditions
at both boundaries. The lepton-number-violating propagators then
satisfy
\begin{gather}
\partial_y Z_p^<(y,y') \big|_{y=0} + m_d Z_p^<(0,y') \,=\, 0, \\
\partial_y Z_p^>(y,y') \big|_{y=L} + m_d Z_p^>(L,y') \,=\, 0, \\[1mm]
H_p^<(0,y') \,=\, 0, \\
H_p^>(L,y') \,=\, 0,
\end{gather}
where the superscripts ${}^<$ and ${}^>$ represent the solutions
for $y < y'$ and $y > y'$, respectively. The Neumann conditions follow
from the integration of (\ref{zp}) over the infinitesimal regions
around $y=0$ and $y=L$, and the continuity of wavefunction. The
derivative of $Z_p$ is jumped at both boundaries due to the existence
of source terms. The solutions with respect to $y$ are found up to
normalizations;
\begin{eqnarray}
Z_p^<(y,y') &=& C_Z^<(y')\big[ q\cosh(qy) - m_d\sinh(qy) \big], \\
Z_p^>(y,y') &=& C_Z^>(y')\big[ q\cosh(qy-qL) - m_d\sinh(qy-qL) \big], \\
H_p^<(y,y') &=& C_H^<(y')\sinh(qy), \\
H_p^>(y,y') &=& C_H^>(y')\sinh(qy-qL).
\label{pprop}
\end{eqnarray}
The functions $C_{Z,H}^{<,>}$ are determined by the conditions
which connect the solutions in two regions, i.e.,
\begin{gather}
Z_p^< \,=\, Z_p^>, \qquad 
\partial_y Z_p^< \,=\, \partial_y Z_p^> - M, \\
H_p^< \,=\, H_p^>, \qquad 
\partial_y H_p^< = \partial_y H_p^> - M,
\end{gather}
at $y=y'$. The discontinuities of the slopes follow from the
integration of (\ref{zp}) and (\ref{hp}) around $y=y'$. The final
result is as follows;

\medskip

\noindent
\fbox{$Z = +1$, ~$Z' = +1$}
\begin{eqnarray}
Z_p^{++}(y,y',m_d,M) &=& \frac{1}{(m_d^2 - q^2)q\sinh(qL)}
\big[ q\cosh(qy_<) - m_d \sinh(qy_<) \big] \nonumber \\[1mm]
&& \qquad\qquad
\times \big[ q\cosh(qy_> - qL) - m_d \sinh(qy_> - qL) \big]M,
\label{propZp} \qquad  \\[2mm]
H_p^{++}(y,y',m_d,M) &=& 
\frac{\sinh(qy_<) \sinh(qy_> - qL)}{q \sinh(qL)}M,
\label{propHp}
\end{eqnarray}
where $y_<$ ($y_>$) stands for the lesser (greater) 
of $y$ and $y'$. The superscript ``++'' is attached to indicate that
the propagators satisfy the boundary conditions $Z=+1$ and $Z'=+1$. 

The mass spectrum in four-dimensional effective theory is extracted
from the poles of these propagators. First, $q^2=m_d^2$ in
(\ref{propZp}) corresponds to the chiral zero mode with the 
mass $M$\@. The other poles, $qL = in\pi$, in both (\ref{propZp}) 
and (\ref{propHp}) give the masses of KK-exited 
states; $m_d^2 + |M|^2 + \big(\frac{n\pi}{L}\big)^2\,$ $(n\geq1)$.

The lepton-number-violating propagators for the other boundary
conditions can be derived in parallel ways to the above:

\medskip

\noindent
\framebox{$Z = +1$, $Z' = -1$}
\begin{eqnarray}
Z_p^{+-}(y,y',m_d,M) &=& 
\frac{\big[ q\cosh(q y_<) - m_d \sinh(q y_<) \big]\sinh(qy_> - qL)}
{q\big[q\cosh(qL) - m_d \sinh(qL) \big] }M,
\label{zppm}  \\[2mm]
H_p^{+-}(y,y',m_d,M) &=& 
\frac{-\sinh(qy_<)\big[ q\cosh(q y_> - qL) + m_d\sinh(q y_> - qL)\big]}
{q\big[q\cosh(qL) - m_d \sinh(qL) \big]}M.  \qquad
\label{hppm}
\end{eqnarray}

\medskip

\noindent
\framebox{$Z = -1$, $Z' = +1$}
\begin{eqnarray}
Z_p^{-+}(y,y',m_d,M) &=& 
\frac{-\sinh(qy_<)\big[ q\cosh(q y_> - qL) - m_d\sinh(q y_> - qL)\big]}
{q\big[q\cosh(qL) + m_d \sinh(qL) \big]}M, \qquad
\\[2mm]
H_p^{-+}(y,y',m_d,M) &=& 
\frac{\big[ q\cosh(q y_<) + m_d \sinh(q y_<) \big]\sinh(qy_> - qL)}
{q\big[q\cosh(qL) + m_d \sinh(qL) \big]}M.
\end{eqnarray}

\medskip

\noindent
\framebox{$Z = -1$, $Z' = -1$}
\begin{eqnarray}
Z_p^{--}(y,y',m_d,M) &=& 
\frac{\sinh(qy_<) \sinh(qy_> - qL)}{q \sinh(qL)}M, \\
H_p^{--}(y,y',m_d,M) &=& \frac{1}{(m_d^2 - q^2)q\sinh(qL)}
\big[ q\cosh(qy_<) + m_d \sinh(qy_<) \big] \nonumber \\[1mm]
&& \qquad\qquad
\times \big[ q\cosh(qy_> - qL) + m_d \sinh(qy_> - qL) \big]M. \qquad
\end{eqnarray}

The last case with $Z = -1$ and $Z' = -1$ gives the same mass spectrum
as that for $Z =+1$ and $Z' =+1$. 
For the other two cases with $Z = \pm 1$ and $Z' = \mp 1$,
the positions of poles are at $p^2 = m_d^2 + |M|^2 + (x_n^{\pm}/L)^2$ 
where $x_n^\pm$ are determined by the 
equations $\tan x^\pm = \pm x^\pm/m_dL$. For small Dirac 
mass $m_d L \ll 1$, the KK indices $x_n^\pm$ approach to
$\big(n - \frac{1}{2}\big)\pi$, which just correspond 
to (\ref{sincos1/2}). In the opposite limit $m_d L \gg 1$, 
the indices become $x_n^\pm \simeq n\pi$ for low-lying modes. 
A special case is $m_d L = 1$ that leads to the eigenvalues $x_n^\pm$; 
\begin{center}
\begin{tabular}{ccccc}\hline\hline
$n$ && $x_n^+/\pi$  && $x_n^-/\pi$ \\\hline
1 && 0 && 0.65  \\
2 && 1.43 && 1.56  \\
3 && 2.46 && 2.54  \\
$\vdots$ && $\vdots$ && $\vdots$ \\ \hline
\end{tabular}
\end{center}
A remark is the appearance of ``zero mode'' $x_n^+ =0$. It is seen
from the propagators (\ref{zppm}) and (\ref{hppm}) that $q=0$ becomes
a pole only if this special relation $m_dL=1$ is satisfied.
A similar pole $x_n^- =0$ appears for $m_d L = -1$.


\end{document}